\documentclass[12pt]{iopart}

\pdfoutput=1
\usepackage{graphicx}
\begin{document}

\title[Entropic measures of individual mobility patterns]{Entropic measures of individual mobility patterns}

\author{Riccardo Gallotti$^\ast$, Armando Bazzani$^\ast$, Mirko Degli Esposti$^\dagger$ and Sandro Rambaldi$^\ast$}

\address{$^\ast$Department of Physics, University of Bologna and INFN - Sezione di Bologna\\
$^\dagger$Department of Mathematics, University of Bologna}

\ead{rgallotti@gmail.com}
\begin{abstract}
Understanding human mobility from a microscopic point of view may represent a fundamental breakthrough
for the development of a statistical physics for cognitive systems and it can shed light on the applicability
of macroscopic statistical laws for social systems. Even if the complexity of individual
behaviors prevents a true microscopic approach, the introduction of mesoscopic models allows
the study of the dynamical properties for the non-stationary states of the considered system. We propose to compute various entropy measures
of the individual mobility patterns obtained from GPS data that record the movements of private vehicles in the Florence district, in order
to point out new features of human mobility related to the use of time and space and to define the dynamical properties of a stochastic model
that could generate similar patterns. Moreover, we can relate the predictability properties of human mobility to the distribution of time passed between two successive trips. Our analysis suggests the existence of a hierarchical structure in the mobility patterns which divides the performed
activities into three different categories, according to the time cost, with different information contents. We show that a Markov process
defined by using the individual mobility network is not able to reproduce this hierarchy, which seems the consequence of different strategies
in the activity choice. Our results could contribute to the development of governance policies for a sustainable mobility in modern cities.
\end{abstract}

\maketitle

\section{Introduction}

Human mobility has recently become a fruitful research field in complexity science, representing a paradigmatic example of statistical systems of cognitive particles.
Thanks to the commercial spreading of the Information and Communication Technologies (ICT), it has been possible to collect large data sets on individual dynamics, offering the possibility of studying this complex system at different scales\cite{vespignani2012}.
As a consequence, important steps have been made toward the characterization of individual mobility patterns: approaches in this direction have been made analyzing data from dollar bills tracking\cite{brockmann2006}, mobile phone\cite{gonzalez2008}\cite{song2010nature}\cite{song2010science}\cite{phithakkitnukoon2010}\cite{bagrow2012}\cite{schneider2013}, subway fare card transactions\cite{roth2011}\cite{hasan2012},
geographic online social networks\cite{noulas2012} and private cars' GPS\cite{schonfelder2004}\cite{bazzani2010}\cite{gallotti2012} data sets.
As a matter of fact, the differing features observed in the various data bases are suggesting different statistical behaviors or explanations.
In most of the papers, the individual mobility is recorded in a indirect way (tracking bills) or in relation to other activities (internet access to social network or use of mobile phone) and the results may be influenced by the complex features the communication habits.
Indeed, the suggested power law distributions for covered distances $p(l)\propto l^{-\gamma}$ ($\gamma\in (1,3]$) and rest times $p(t)\propto t^{-\beta}$ ($\beta\in (1,2]$)
seem not to hold in urban contexts, where the mobility is limited by the energy consumption for the physical activity of daily travel\cite{kolbl2003} and is therefore dominated by short trips\cite{noulas2012}\cite{bazzani2010}\cite{liang2012}.
At the beginning of every new day, people wake up knowing that a series of tasks have to be carried out.
Necessarily, they must satisfy the physiological need of eating, which represents a periodical constraint in the daily routine,
and sleeping, which forces the circadian rhythm and thus imposes an end to the chain of performed activities.
Besides, they have to perform duties, usually the working related ones, that are precisely scheduled and others,
as shopping, social and leisure activities, that can be done in any moment during the free time.
Changes of plan during the day are always possible: a planned meeting may be postponed, a new activity may be picked instead
of another or something can be done just because some spare time is available.
The accomplishment of the daily duties is realized traveling through the locations where each particular task has to be performed,
and it is clearly the primary cause of human mobility. Therefore, understanding the structure of the daily activity pattern is a crucial step in the development of a model for human mobility, thereby leading to applications in fields like urban planning, traffic management, or epidemic spreading.
But it can also contribute to a formulation if a statistical mechanics theory for complex systems. This theoretical goal
requires three fundamental steps:
\begin{enumerate}
\item to discover generic macroscopic behaviors (i.e. statistical laws) describing average properties of human mobility;
\item to characterize the individuals' microscopic dynamics and its relation with the macroscopic properties;
\item to study the transient states driven by external changes and the existence of critical phenomena (i.e. phase transitions)
due to interactions among individuals.
\end{enumerate}
The first item has been considered by several papers and various statistical laws for human mobility have been proposed\cite{brockmann2006}\cite{gonzalez2008}\cite{song2010nature}\cite{bazzani2010}\cite{gallotti2012}
trying to point out the common features and the differences with the physical models. These discoveries of statistical laws imply also the definition of relevant macroscopic observables representing the state of the system.
More recently, the scientific community has also faced the second item in order to understand the cognitive aspects at the base
of individual mobility (i.e. mobility strategies) and to discover the primary causes of the macroscopic behaviors\cite{zhou2008}. On the contrary, the last item seems to be at present time beyond the possibility of the Complexity Science.

Many questions are still open, in particular concerning the applicability of a statistical approach to the individual mobility. 
The goal of this paper is to study the dynamical features of individual mobility that can be related to entropy measures
and to contribute in the understanding of their relationship with the empirical statistical results on the use of space and time in vehicular mobility. In particular, we refer to the Benford's law $p(t)\propto t^{-\beta}$ $\beta\simeq 1$ that seems to well describe the  distribution of the time $t$ between two successive trips when $t\le 4$h and a power law $p(k)\propto k^{-\alpha}$ with $1<\alpha<2$ for the ranking distribution of the individuals' most visited locations\cite{song2010nature}\cite{gallotti2012} and we discuss how the structure of the individual mobility patterns can be at the base of such distribution laws. Mobility patterns entropy has been considered other papers to study the predictability properties
of human mobility\cite{song2010science} and to suggest a classification of individual heterogeneity in order to define mesoscopic models
that reproduces the macroscopic empirical laws. These results are based on mobile phone tracking datasets and they
mainly concern the time mobility patterns of individuals (the time scale is $\simeq 1$h)which
are dominated by long term activities (stay at home or work) related to the circadian rhythm. Focusing on urban mobility
the previous analysis could hide the relevance of \textit{asystematic} (random) mobility which characterizes many short activities
that each individual performs in modern cities. In this paper we take advantage from a data base on single vehicle
mobility where a sampling of each trajectory is recoded using a GPS system\cite{octo} in whole Italy. Approximately $2\%$ of the entire vehicle population
is monitored for insurance reasons and time, position and covered distance are recorded every 2 km or 30 sec. Moreover a signal is
also recorded each time the engine is switch on or off. Even if one have no information on the vehicle sample for privacy reason,
this data base offers a unique opportunity to study individual mobility at fine spatial scale on large urban areas and using
long time series. More precisely in the sequel we analyze the GPS data base recorded in Florence during March 2008.
In order to highlight the different roles that different duration activities have to define
the individual mobility patterns, we have chosen to use entropy as the key quantity for our study
and we have introduces methods of analysis different from those used in\cite{song2010science}.
Indeed, for our purposes entropy is a good observable, because it resumes the information present in a
sequence of characters (given by the statistical properties and the correlations due to the personal habits) in a real number.
In particular, as the entropy per character in principle does not depend on the length of the sequence,
we can then easily compare patterns of different length representing the mobility of different individuals.
\par\noindent
The paper is organized as follows:
\begin{enumerate}
\item in the first section we discuss the main features of the GPS database used to detect the individual mobility patterns;
\item in the second section we define the entropy measures considered in the paper and their properties;
\item in the third section we study the time regularity of individual mobility and we analyze the
entropic properties of temporal and spatial patterns, suggesting possible explanations for the underlying dynamics;
\item in the last section we propose a simple Markov model on a mobility network, which simulates the entropic properties of
the empirical mobility patterns.

\end{enumerate}

\section{Data}

In Italy, over the 2\% of the private vehicle population is equipped with a GPS system that tracks their path on the road network for insurance reasons.
A database is recorded with geographical coordinates, time, velocity and path length of
these individual trajectories sampled every 2 km, or alternatively 30 seconds in highways. In addition, special signals are registered
when the engine is switched on and off. If the quality of the satellite signal is good, the time precision of the recorded data is practically
perfect whereas the space precision is of the order of 10 meters. A filtering procedure has been applied to exclude positioning errors due to low signal.
This database allows to obtain extremely detailed information on human mobility since one
has the possibility of directly following the people movements.
In fact, we can easily and precisely define a trip as the displacement between two locations where the engine has been turned off more than
a given time threshold (we choose 5 minutes).
In this paper we use data collected during the whole month of March 2008 in the municipality area of Florence.
We study the urban mobility points of 2360 individual among $\approx$32.000 users. The selected individuals respect two criteria.
The first criterion is that these vehicles have been used on each of the 20
working days of the considered month: we want to exclude occasional users and focus on the city users. The second criterion is that there
should not have been any major loss of signal. In fact, the GPS device loses the satellite signal frequently at starting points of the trajectories or
when vehicles are parked inside a buildings. We have then excluded all individuals who have had one or more trips where the origin lies in a
different location then the preceding destination.
We assume that every time the engine was off for more than 5 minutes this stop can be associated to an activity performed near the parking place.
From the cloud of all the parking points of a particular individual the different activity locations have been identified with a gravitational
clustering algorithm. In this clustering mechanism, a maximum acceptable distance of 400m\cite{benenson2008} between the true destination
and the parking place has been assumed. Once the locations are identified, at each of them is associated an identification number that,
in analogy with the studies of the entropy of texts, we call {\em character}.
Finally, we have isolated two types of mobility patterns for each individual. The first is the time pattern, where the month has been divided
in equal time intervals of length $\Delta t$ and at each time interval we have associated an activity (in this case, traveling is considered
as an extra activity and therefore is identified by a specific character). As it is possible that many activities have been performed in the same
time frame, one activity has been randomly chosen in case of conflict regardless the activities relative duration. This procedure is the same implemented
in the paper\cite{song2010science}, allowing to compare results with precedent studies. We remark that time patterns
built with time interval $\Delta t$ may neglect activities shorter than $\Delta t$, which disappear from the further analysis.
The second type of pattern is the jump pattern that consists in the sequence of visited locations. According to our choice, a location is
introduced in the individual mobility when the engine is switched off more than a time interval
$\Delta t\geq 5$ minutes. Therefore activities that require a time shorter than $\Delta t$ are excluded from the sequences.
If, after this exclusion, two or more identical character appear one aside the other, they collapse in an unique copy of the same character,
so any repetition of the same character is neglected.
Clearly, these two types of patterns carry different information. In the first case the main contribution to the mobility patterns is due to
the long term activities. In the second case one focuses the attention on the pattern distribution of different activities.

\section{Entropy Measures}

Following\cite{song2010science}, we consider three different  information entropy measures for a mobility pattern: the information entropy $S$,
the temporal-uncorrelated (Shannon) entropy $S_u = - \sum_{j=1}^N p_j \log_2 p_j$ and the random entropy $S_r = \log_2 N$, where $p_j$ is probability
of finding the character $j$ in our sequence and $N$ the number of distinct characters in the sequence. $S_r$ represents a maximal value of the entropy,
because for representing the mobility pattern it is necessary at least a number of characters equal to the total number of visited activities.
$S_u$ is a better measure of the entropy, as it considers also the uneven frequencies of the characters in the sequence, while still ignoring the
possible compression due to the relative position of the characters, that is taken into account by the information entropy.
These three values are clearly bound by the relationship $S \leq S_u \leq S_r$.
The measure of the information entropy per character $S$ has been estimated using a Lempel-Ziv (LZ) algorithm estimator. This algorithm searches for repeated
sequences of characters that may be exploited for the data-compression of the pattern. More precisely, for a sequence of length $n$ the estimated value of entropy is:
\begin{equation}
S =  \left ( \frac{\sum_{i=2}^n l_i}{n\log_2 n} \right )^{-1}
\end{equation}
where $l_i$ is the length of the shortest string starting at position $i$ that does not appear in the part of sequence up to position $i-1$ (included).
The goodness of this estimate raises with the length $n$ of the sequences and decreases with the broadening of the alphabet size $N$.
The information entropy is related to the concept of predictability $\Pi$, defined as the rate of correct
predictions about the value of a character in the sequence knowing all the precedent characters.
The more informative is a sequence the more difficult is to predict how it may continue: a low entropy is related to a high predictability
and vice-versa. An upper bound $\Pi^m$ to the predictability of a sequence can be computed as a function of the entropy $S$ by the inversion of the formula:
\begin{equation}
-\Pi^m\log_2\Pi^m - (1-\Pi^m)\log_2(1-\Pi^m)+(1-\Pi^m)\log_2(N-1) = S
\label{eqPred}
\end{equation}
This last equation is a consequence of the Fano's inequality\cite{cover} and the inversion is possible when $N$ and $\Pi^m$ are not too small.

\section{Time Regularity}

\begin{figure}[htc!]
\centerline{\includegraphics[angle=0,
width=.7\textwidth]{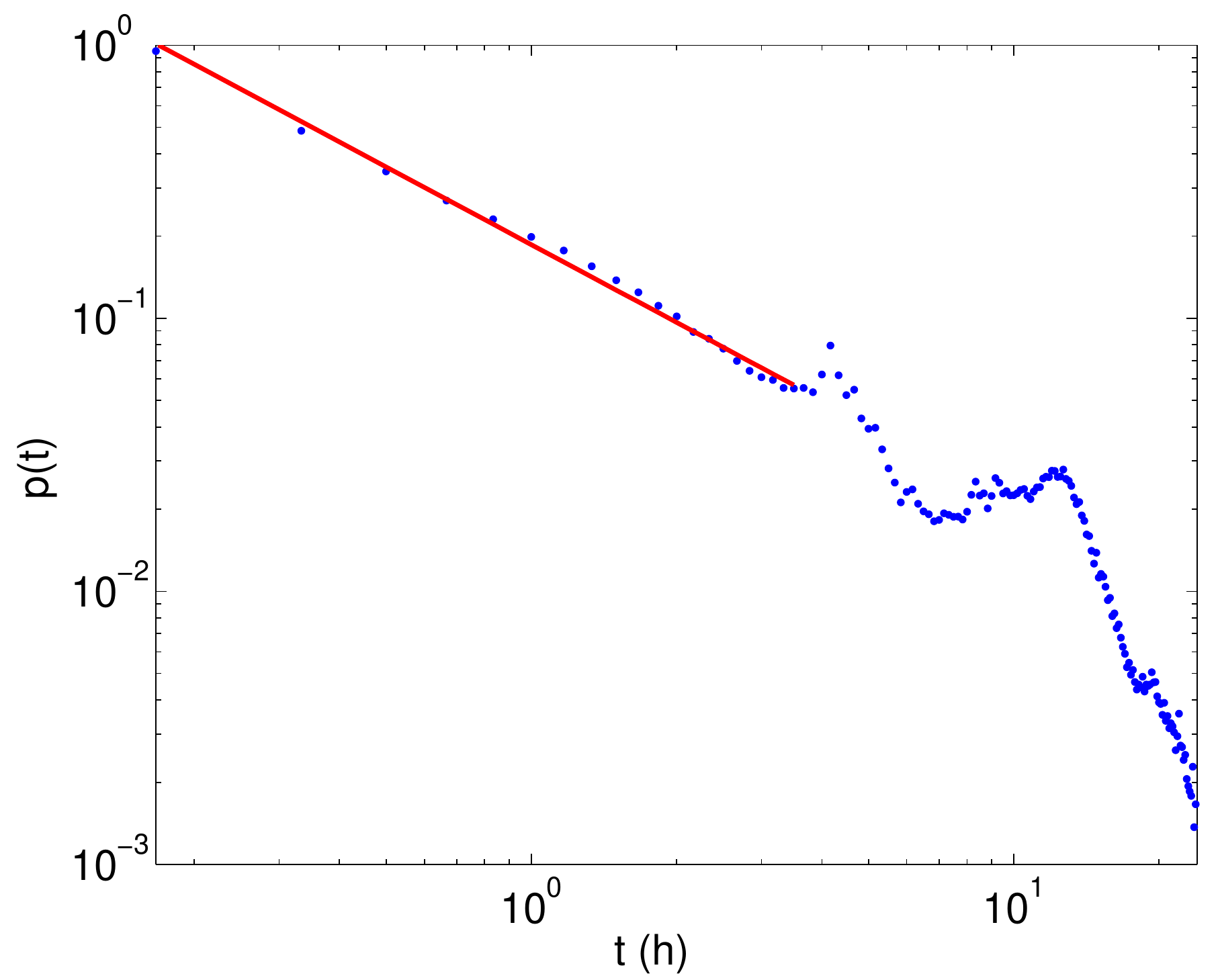}}
\caption{The downtime
distribution (blue dots) for the considered users in the area of
Florence during March 2008. The downtimes are computed
considering the vehicle stops longer than 5 minutes. The red line
suggests an interpolation with a Benford's law $p(t)\propto
t^{-\beta}$ $\beta=.95\pm .04$ for activities shorter than 4h.}
\label{benford}
\end{figure}

\begin{figure}[htc!]
\centerline{\includegraphics[angle=0, width=.9\textwidth]{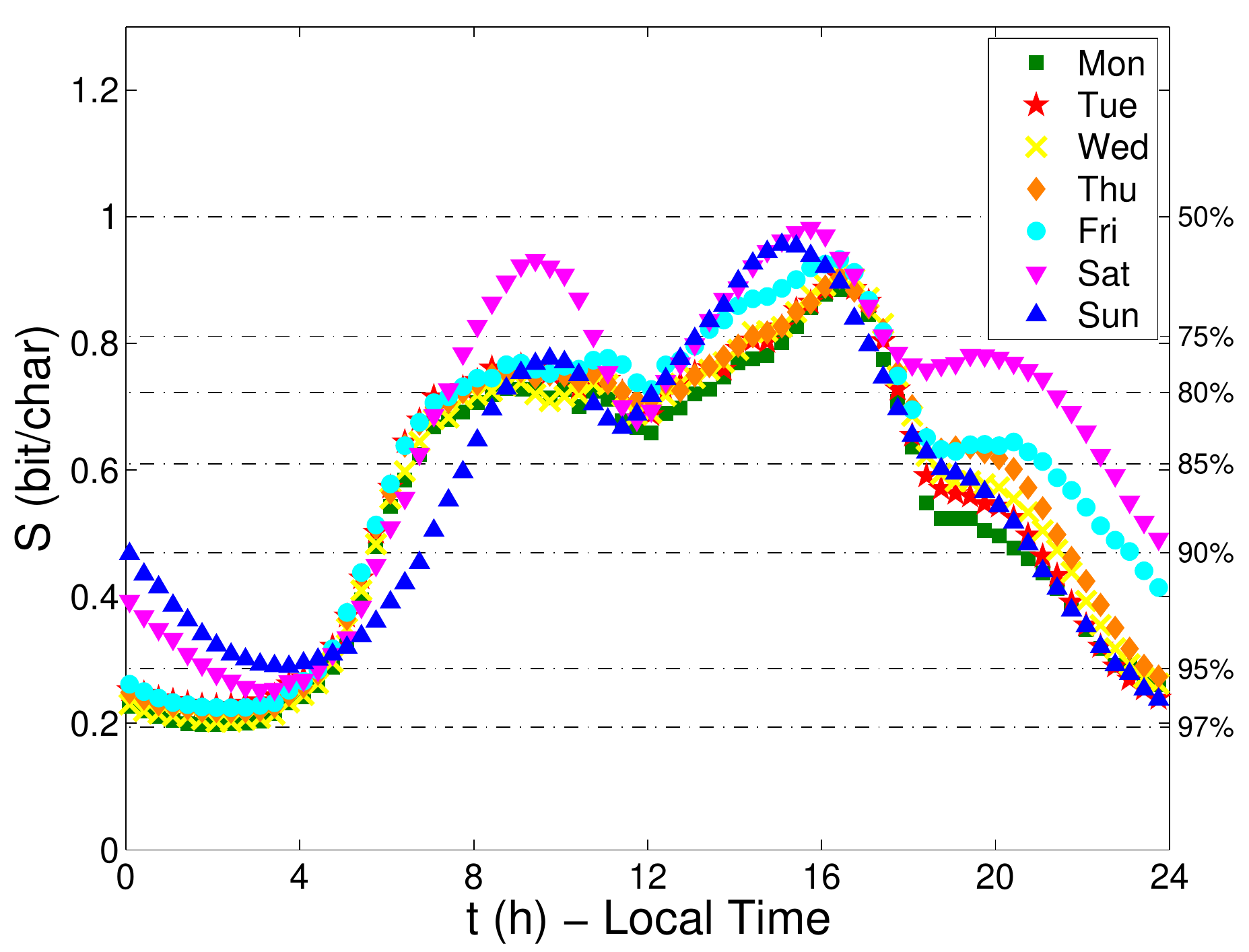}}
\caption{The average Regularity Entropy $S_u(t_i)$ measures the information lying in the distribution of the probabilities of finding a
person in a particular place in a given time frame. Every day of the week has a characteristic hour-dependency.
Monday: green squares; Tuesday: red stars; Wednesday: yellow crosses; Thursday: orange diamonds; Friday: cyan circles; Saturday:
magenta down triangles; Sunday: blue up triangles. With equation (\ref{eqPred}), we can link this entropic measure with the upper bound to
predictability. The dot-dash lines represent the related values of $\Pi^m(S)$, calculated for $N=2$.
}
\label{regEnt}
\end{figure}

We have considered 28 days, so that we have 4 consistent record for each day of the week. At this purpose, we excluded two public holidays and a Sunday when the daylight saving time has been introduced. Because of the daylight saving time, we have had to compensated one hour for the 2 last days of the month days following that Sunday.
The downtime distribution plotted in the figure \ref{benford} is well interpolated by a
Benford's law $p(t)\propto t^{-\beta}$ with $\beta\simeq 1$\cite{bazzani2010} (red line) up to times of few hours,
whereas we clearly see three peaks corresponding to 4h,
8h and 9h (partial and full work time) and a broader one around 12h (rest time at home)\footnote{This value may be explained by a
logarithmic time perception, or alternatively if the daily schedule is created progressively,
with in the activities duration limited by the free time left in the timetable.}.
This fat tail distribution points out that the use of time during mobility is
dominated by the longer term activities, which is reasonable to associate to a daily origin-destination mobility.
For each time frame $[t_i,t_i+\Delta t]$ with $\Delta t=20$ minutes, representing a moment in the day, we have counted how many times the drivers were in their most visited location for that moment.
We use this  for each user to compute the \textit{regularity} $R(i)$, i.e. probability of finding the user in his most visited location at the time frame $i$,
whose distribution among the users is within the $79\pm9\%$ confidence interval, which is consistent with
the result reported in\cite{song2010science} based on mobile phone data.
However, considering only the most visited location does not take into account the complete information contained in the
visited locations signal at each time frame $i$.
An entropic approach can here grant a most comprehensive analysis. Thus we have calculated the Shannon entropy
$$
S_u(t_i) = - \sum_{j=1}^{N} p_j(t_i) \log_2 p_j(t_i)
$$
for the probability distribution $p_j(i)$ to find a driver at the location $j$ at each time frame $i$. The average values of $S_u(t_i)$ across all users for the different days of the week is shown in fig.\ref{regEnt}. We remark how every day has a peak of dispersion
in the afternoon where one tends to perform leisure or social activities and has its minimum value in the late night (most people are at home).
The entropy value in the morning during working days corresponds to a predictability between $75\%-80\%$ which is consistent with
the predictability of the most visited locations: this may be an indication that people mobility is related to working activities in the morning.

As could have been expected, Saturday shows the greatest variability with an average entropy $\langle S_u(t_i)\rangle$ of 0.66 bit/char, $\approx 15\%$ higher of the average entropy of working days. The working days are substantially similar within the standard error ($\pm$0.01 bit/char), if we exclude a growing tendency to spend time in unexpected places during the evenings (and in Friday afternoon). As a consequence of that, the average entropies are slightly but steadily growing (Monday: 0.53 bit/char, Tuesday and Wednesday: 0.56 bit/char, Thursday: 0.58 bit/char, Friday: 0.61 bit/char). On Sunday, the mobility appears to have a late beginning and thus, despite high values of entropy in the afternoon, the average value of 0.57 bit/char is similar to those of working days. The relative standard deviations, representing how the value of $S_u(t_i)$ changes during the day, have consistent values across all days within the range [0.22,0.23] bit/char.

\section{Pattern Analysis}

\begin{figure}[htc!]
\centerline{\includegraphics[angle=0, width=.45\textwidth]{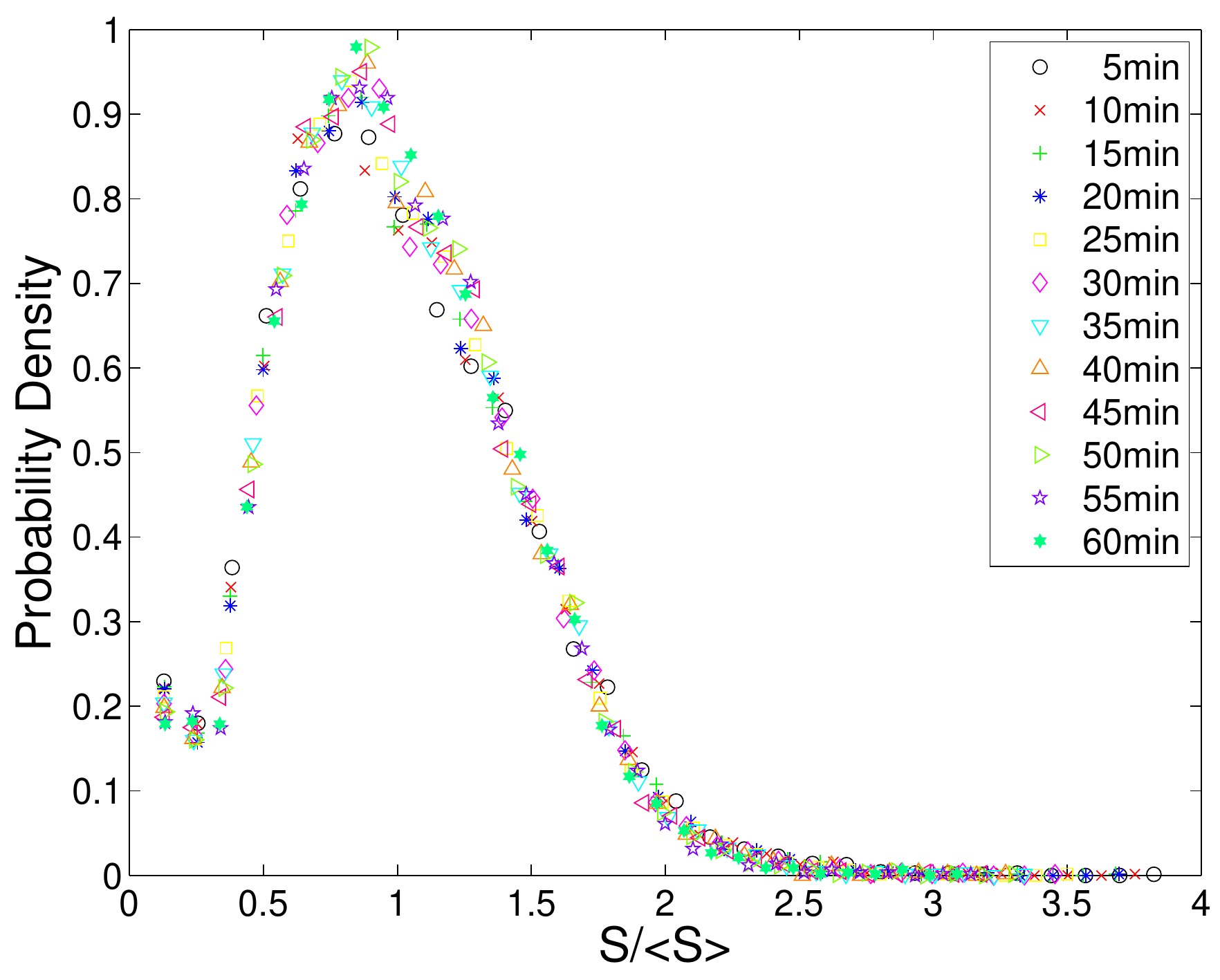}\quad\includegraphics[angle=0, width=.45\textwidth]{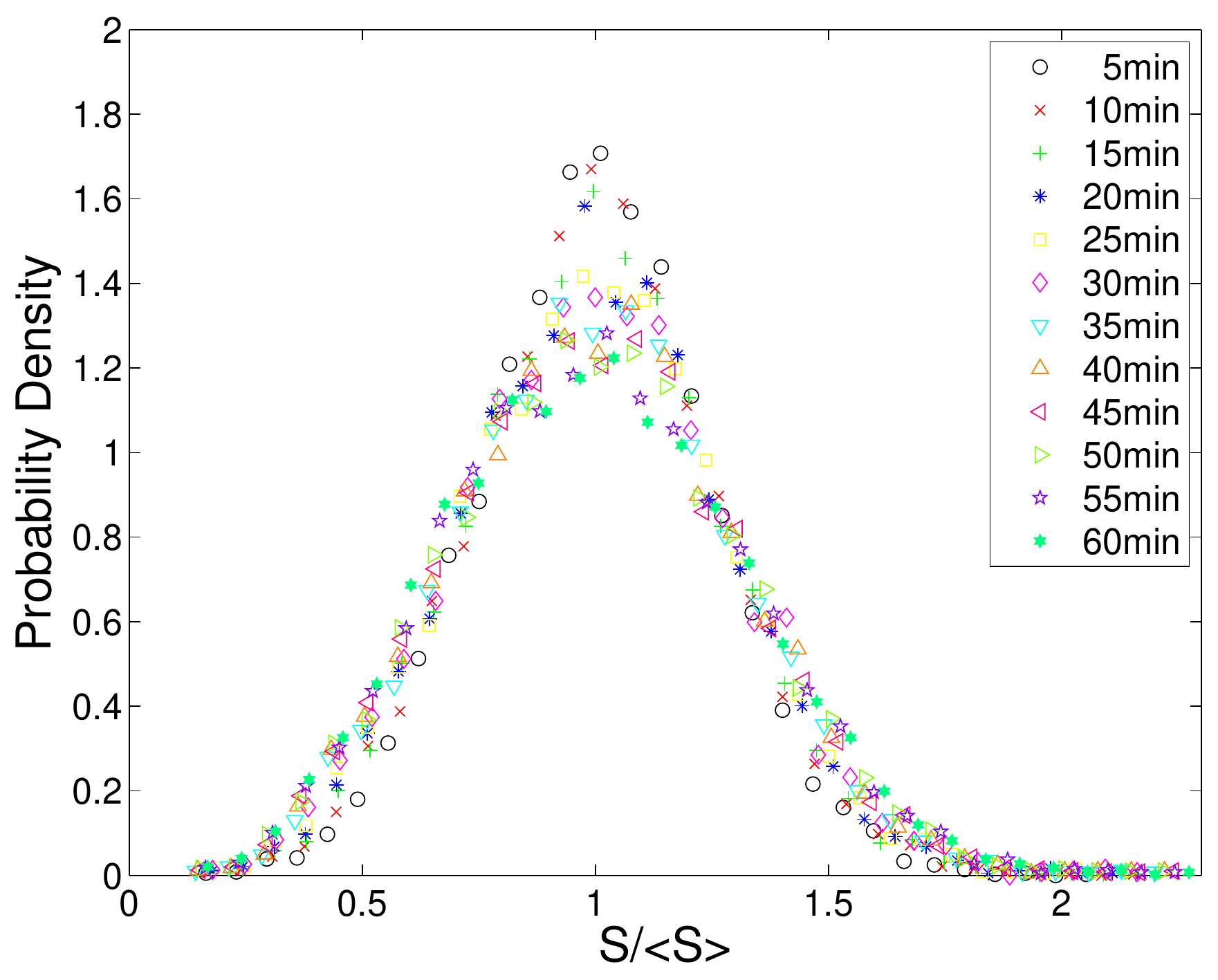}}
\caption{\textbf{(Left picture)} The rescaled distributions of the Time Patterns Entropies $S$ with different time-frames
$\Delta t$ are perfectly superposed, suggesting that all the dependency upon $\Delta t$ lies in the average value $\langle S\rangle$.
\textbf{(Right picture)} Similarly, the rescaled distributions of the Jump Patterns Entropies $S$ with different threshold $\Delta t$
are superposed and also in this case all the dependency upon the threshold lies in the average values of the entropy.}
\label{universal}
\end{figure}

\begin{figure}[htc!]
\centerline{\includegraphics[angle=0, width=.45\textwidth]{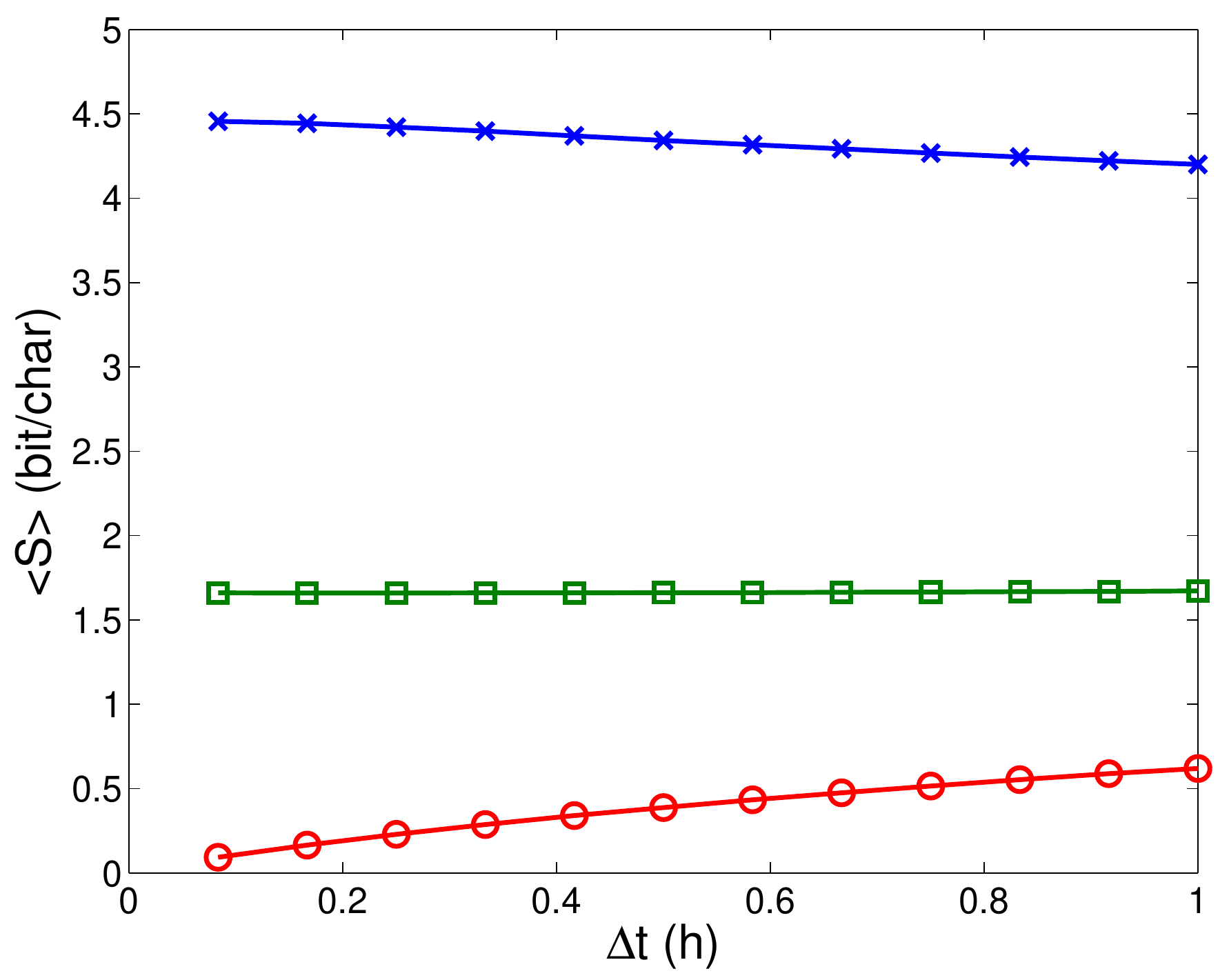}\quad
\includegraphics[angle=0, width=.45\textwidth]{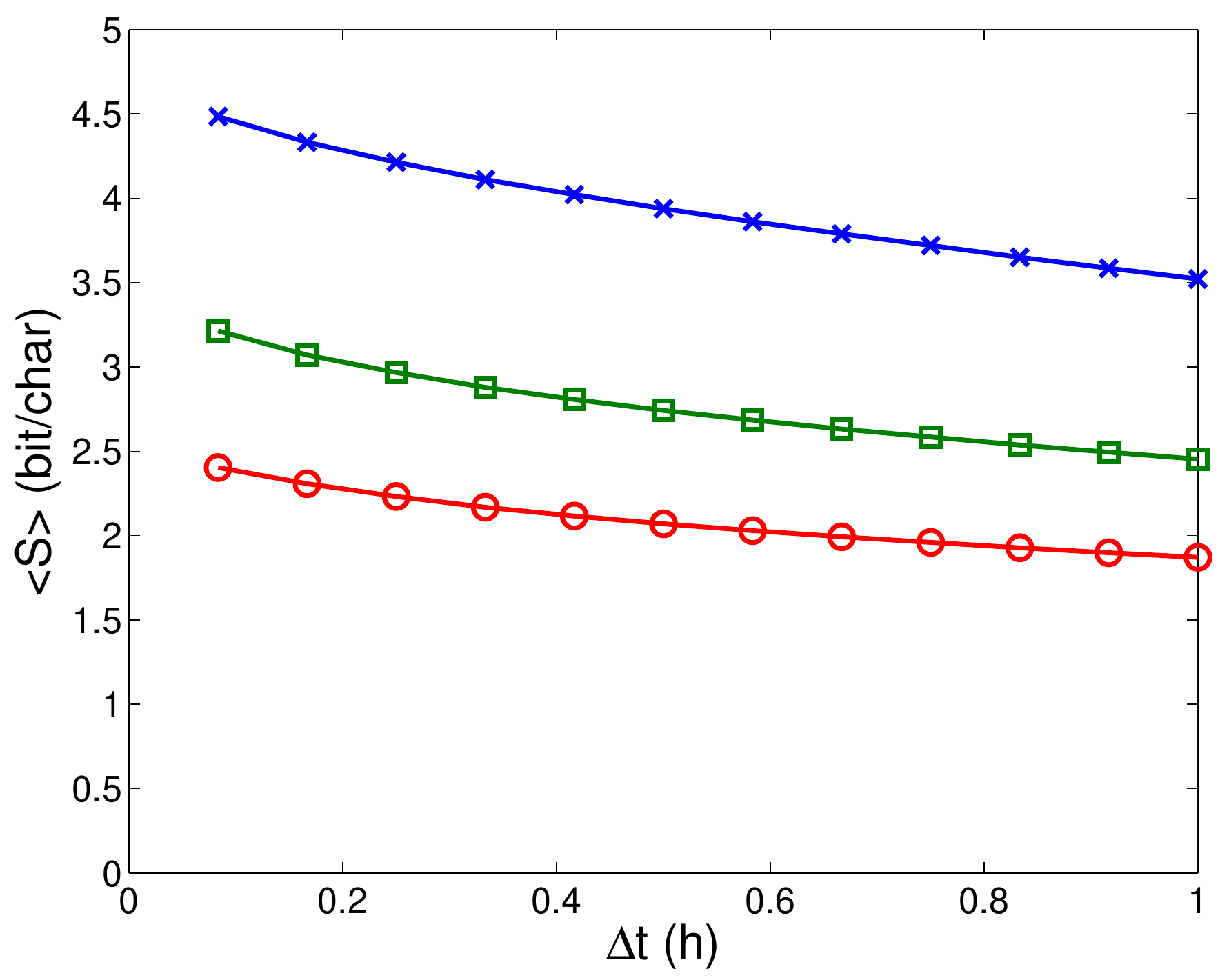}}
\caption{\textbf{(Left picture)}The dependency upon the timeframe size $\Delta t$ of the average values of the Time Patterns Entropies.
$\langle S_r\rangle$ (blu crosses) grows for small values of $\Delta t$, due to the inclusion of new locations, while $\langle S_u\rangle$
(green squares) is constant, indicating that the frequency distribution of the different characters in the sequences $p_j$ is not dependent on the size of the timeframe.
$\langle S\rangle$ (red circles) drops for small $\Delta t$. \textbf{(Right picture)}The dependency upon the threshold $\Delta t$
of the average values of the jump Patterns Entropies. In this case all entropy averages $\langle S_r\rangle$ (blu crosses), $\langle S_u\rangle$ (green squares)
and $\langle S\rangle$ (red circles) grow for small values of $\Delta t$.}
\label{tpent}
\end{figure}

Being our data extremely precise in time, we can afford to create time patterns and jump patterns with short time frame values $\Delta t\geq 5$min.
This extension could be important to characterize the dynamical properties of individual mobility: e.g. are the short activities
chosen randomly whereas long term activities are related to the individual habits according to the circadian rhythm? May the short
activities justify the applicability of statistical physics distribution to human mobility?
First we consider the dependence of the entropy distribution from the time frame threshold $\Delta t$
used to define the mobility patterns. In the figure \ref{universal}, we report the distribution of the normalized entropy $S/\langle S \rangle$
using different time frames both for the time patterns and for the jump patterns.
The results, regarding time frames ranging from 5 to 60 minutes, suggest the existence of an universal probability distribution $p(x)$
which represents the entropy distribution of the mobility patterns according to
\begin{equation}
P(S; \Delta t)=\frac{1}{\langle S\rangle(\Delta t)}p\left (\frac{S}{\langle S\rangle(\Delta t)}\right )
\label{distri}
\end{equation}
Assumed that decreasing the threshold $\Delta t$ we add new activities in the patterns and an increase of the Entropy value
is expected, the dynamical meaning of eq.(\ref{distri}) is that the structure of the mobility patterns does not depend on the activity
duration and we expect a self-similarity in the patterns at different time scales\cite{proekt2012}.
Since the only significative value representing the dependence of the distribution $p(S)$ on $\Delta t$ is the mean value $\langle S \rangle$, in the following
we will focus on the study of the mean values of the entropy. In fig.\ref{tpent} we plot the dependence of average entropy values from
the time-frame size for the three entropy measures considered. The left picture concerns the time mobility pattern and it is remarkable
to observe that the average Shannon entropy $\langle S_u\rangle$ is independent from $\Delta t$, denoting that the relative frequencies
of the time patterns remain almost constant. Indeed this is a consequence of the Benford's laws $p(t)\propto 1/t$ which implies that frequency
of the activities of duration $t$ is $\simeq t/\Delta t$ with respect to a total length $T/\Delta t$ so that the relative frequency
is $t/T$. On the contrary, the information entropy decreases due to weight of the long term activities that increases the persistence in the patterns:
i.e. the probability of continuing an activity up to $t+\Delta t$ given that the activity $t$ has been carried out for a time $t$.
A rough estimate of the persistence based on the Benford's law gives
$$
\pi(t+\Delta t/t)\propto 1-c\ln \left (1+\frac{\Delta t}{t}\right )
$$
where $c$ is a suitable constant so that $\pi(t+\Delta t/t)\to 1$ when $t$ increases. Therefore we expect that
the entropy $S$ tends to zero for short time frames.
Indeed, the most part of the new characters introduced shortening $\Delta t$ are only prolonging repetitions of the same character
representing long stops. Even if we are introducing new information regarding shorter activities in our analysis, the analyzed sequences
will have more and more longer series of iterated characters, which can be easily compressed and therefore lower the value of entropy per character.
\par\noindent
This argument does not apply to the jump patterns where the increase of the average Shannon entropy with $\Delta t$
is direct indication that the number of activities in the pattern is increasing (see fig.\ref{tpent}).
We also remark as the values of $S_u$ and $S$ are greater in the
jump patterns, as we have neglected all the repetitions that were dominating the distribution of character frequencies and are easily compressible.
Moreover, the Shannon entropy measure $S_u$ can be related to
the ranking distribution of most visited locations $p(k)\propto k^{-\alpha}$ that has been
proposed by various authors\cite{song2010nature}\cite{gallotti2012} in mobility data. A simple computation gives
$$
S_u=-\sum_k p(k)\log_2 p(k)\simeq -\frac{\alpha-1}{\ln 2}\int_1^\infty k^{-\alpha}\ln (\alpha-1)k^{-\alpha} dk=
\frac{1}{\ln 2}\left (\frac{\alpha}{\alpha-1}-\ln (\alpha-1)\right )
$$
A value $\alpha\simeq 1.7$ is in accord with the GPS observations in Florence\cite{bazzani2010} and the previous crude estimate gives $S_u\simeq 4$
to be compared with $S_u\simeq 3.4$ obtained by a direct calculation (see fig.\ref{tpent}).

\subsection{Time Patterns }

\begin{figure}[htc!]
\centerline{\includegraphics[angle=0, width=.5\textwidth]{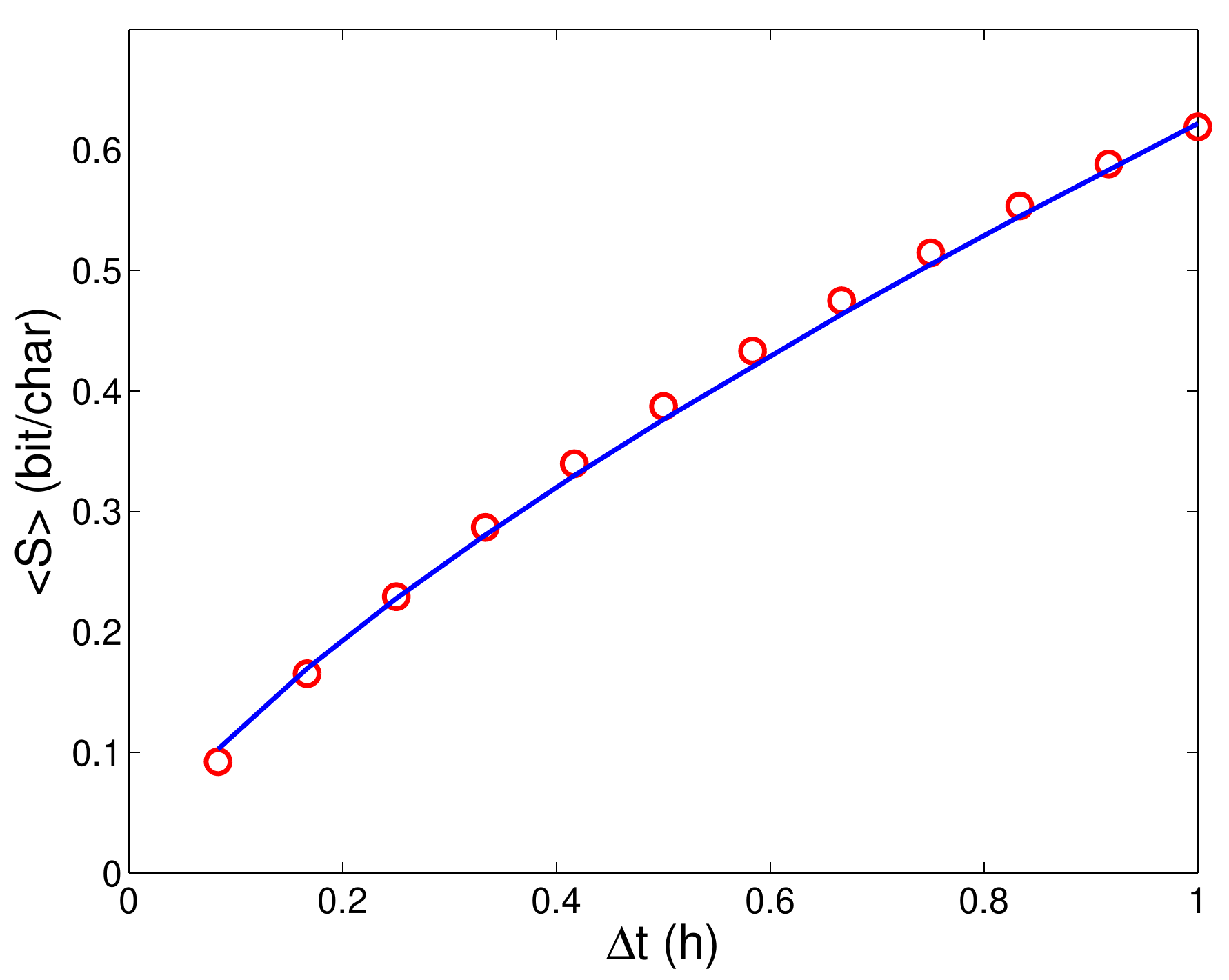}}
\caption{The scaling of $\langle S\rangle$ can be explained measuring the entropy of Monte Carlo simulated sequences that share the same Benford's law
distribution of the length of repeated characters. Thus, the fact that the entropy tends to zero for small time frames is only due to the progressive growth of
length of the sequences of repetition of the same character that describes the same activity.}
\label{scale}
\end{figure}

The threshold value $\Delta t=60$ minutes represents the same experimental conditions considered in the paper\cite{song2010science} for
mobile phone users' mobility. The reported results show an upper limit to the predictability of $93\%$ that is in a
remarkable accord with value of $\Pi^m=(92\pm3)\%$ that we can derive from our measures of $S$. The validation of this result on an independent,
and rather different, data set grants that mobility patterns obtained from both phone calls and private cars' GPS  data sets are a good representation
of the human mobility with a temporal scale of one hour.  But, using a time frame of this size, we are keeping out a great part of the mobility,
as the $41\%$ of the stops during urban mobility are shorter than one hour. As previously remarked, the information entropy tends to zero
when $\Delta t\to 0$ (see fig.\ref{tpent}).
A numerical interpolation of the empirical data suggests that $S$ follows a scaling law $S(\Delta t) \propto \Delta t^{\beta}$, where $\beta = 0.75\pm0.03$.
We will show that this scaling law is a consequence of the Benford's law for downtime distribution. We have generated synthetic sequences
constituted only by two characters (0 or 1). This choice permit us to reduce the errors of the LZ estimator, that converges slower to the
real value of $S$ for larger numbers of characters in the used alphabet. A random character is picked and is repeated $r$ times, with $r$
distributed as $p(r)\propto r^{-1}$. Then another random character is picked and repeated, and so forth.
Sequences of these kind, which total length equal to the total number of minutes in a month, have been generated and shorter sequences,
corresponding to the different time frame $\Delta t$, have been derived from the original sequences. The information entropy $S_s$ of these shorter
sequences have been estimated with the LZ algorithm and it turns out that the results are well interpolated by a scaling low
$S_s(\Delta t) \propto \Delta t^{\beta_s}$ where $\beta_s = 0.73\pm 0.03$ (see fig.\ref{scale}).

As a consequence, time patterns are not convenient to study the human mobility related to short activities unless
one scales the spatial resolution together with the time resolution to avoid these repeated characters.
Unluckily, our data do not permit this fine analysis,
and therefore we cannot extract any activity time related feature from our time patterns.
However, we introduce a different approach to this issue by considering the jumps patterns.

\subsection{Jump patterns}

We have seen that the information entropy of the time patterns is unsuitable for describing the structure of the human
mobility at short time scale since the statistical distribution of the stop duration plays the leading role in determining
the value of $S$. In order to point out the structure of the transitions from one location to another,
we have analyzed the jump patterns, where only movements are considered.

\begin{figure}[htc!]
\centerline{\includegraphics[angle=0, width=.5\textwidth]{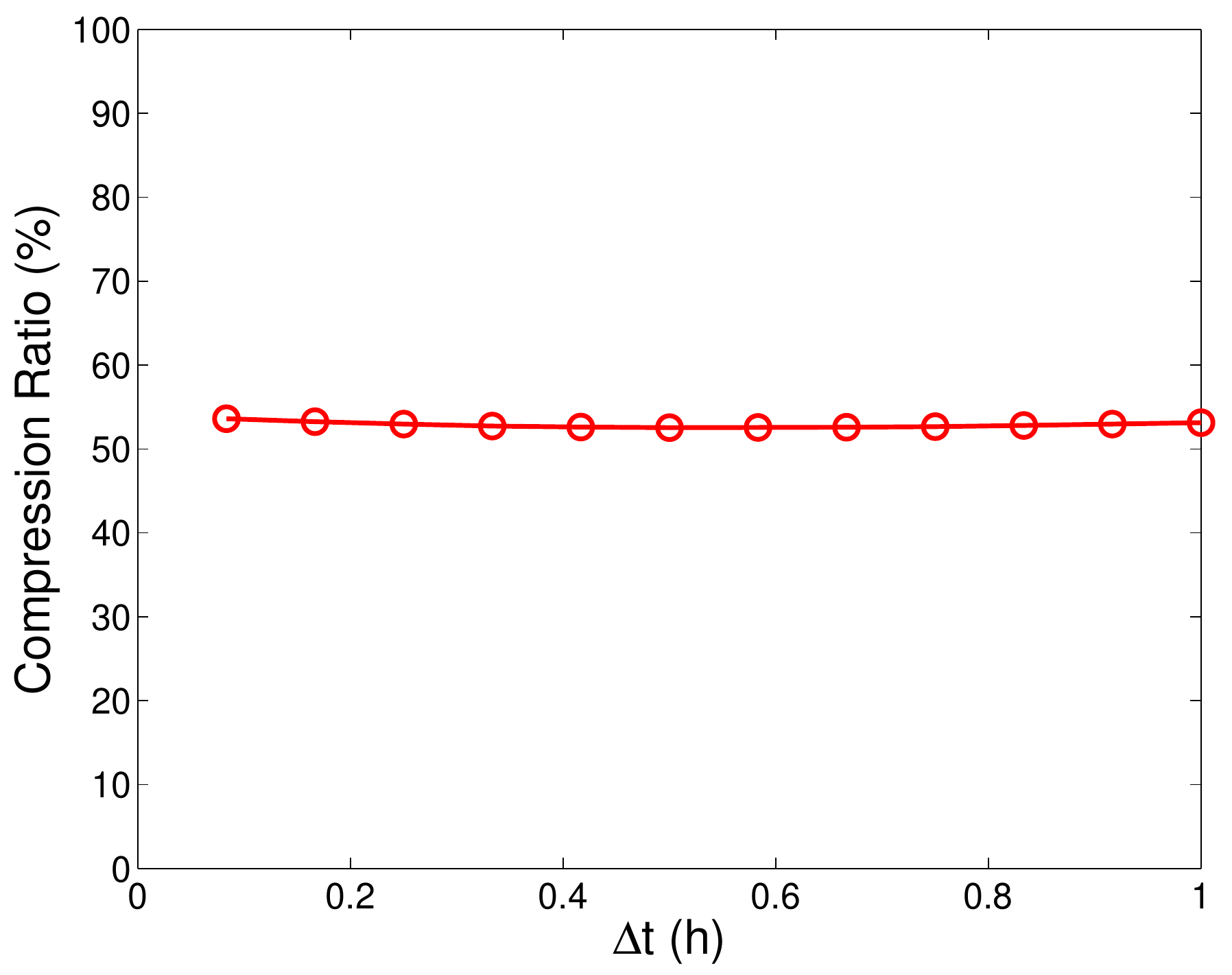}}
\caption{ \textbf{(Right picture)} The compression ratio $\frac{\langle S\rangle}{\langle S_r\rangle}$ has no dependency on the threshold value.
This suggests a time invariant structure of the correlations including activities shorter than one hour.}
\label{tpentJ}
\end{figure}

We consider again values of $\Delta t$ ranging from 5 to 60 minutes, that represent a time threshold under
which stops are excluded from the pattern. Larger thresholds can hardly be considered as it would produce strings too short to be conveniently
analyzed with the LZ algorithm.  Also in this case, we focus the attention on the dependence of entropy mean value from the time threshold
since the distribution of the normalized entropy measures is independent from $\Delta t$ (see fig.\ref{universal}).
From fig.\ref{tpent} we observe that shortening the time frame and thus introducing new characters, all the three measures of entropy raise.
This is a straightforward consequence since introducing new characters we introduce new information. Although, we remark that the change in $S$ is not
great\footnote{Translated in maximum predictability through the inversion of equation (\ref{eqPred}), the values vary from 66\% to 71\%.},
as common sense would suggest since the shorter activities, being free from organizational constraints, might appear more
randomly within the mobility pattern than the longer ones. We would have expected a steeper increase of the entropy caused by the emergence of
asystematic behaviors associated to shorter stops. Not only this is not the case, but if we compute the compressibility
ratio $S/S_r$ it turns out that it has a very weak dependence on $\Delta t$(fig.\ref{tpentJ}).
The compression ratio is the measure of the maximum possible compression that can performed on the string (it is equivalent to the ratio between the size
of a zipped file and the size of the original file): for the considered values of $\Delta t$ i our analysis, its value lies between 52.6\% and 53.6\% .
This fact clearly indicates that also the short activities within the range 5-10 minutes are as compressible
as long ones within the range 50-60 minutes, since they have the same chance of being part of a repeated sequence that the LZ estimator uses for the compression.
As a consequence, we conjecture that the mobility patterns are somehow invariant with respect of the rescaling of activity time,
within the analyzed range (5-60 minutes). A similar observation has been considered in the paper\cite{proekt2012}.
\begin{figure}[htc!]
\centerline{\includegraphics[angle=0, width=.7\textwidth]{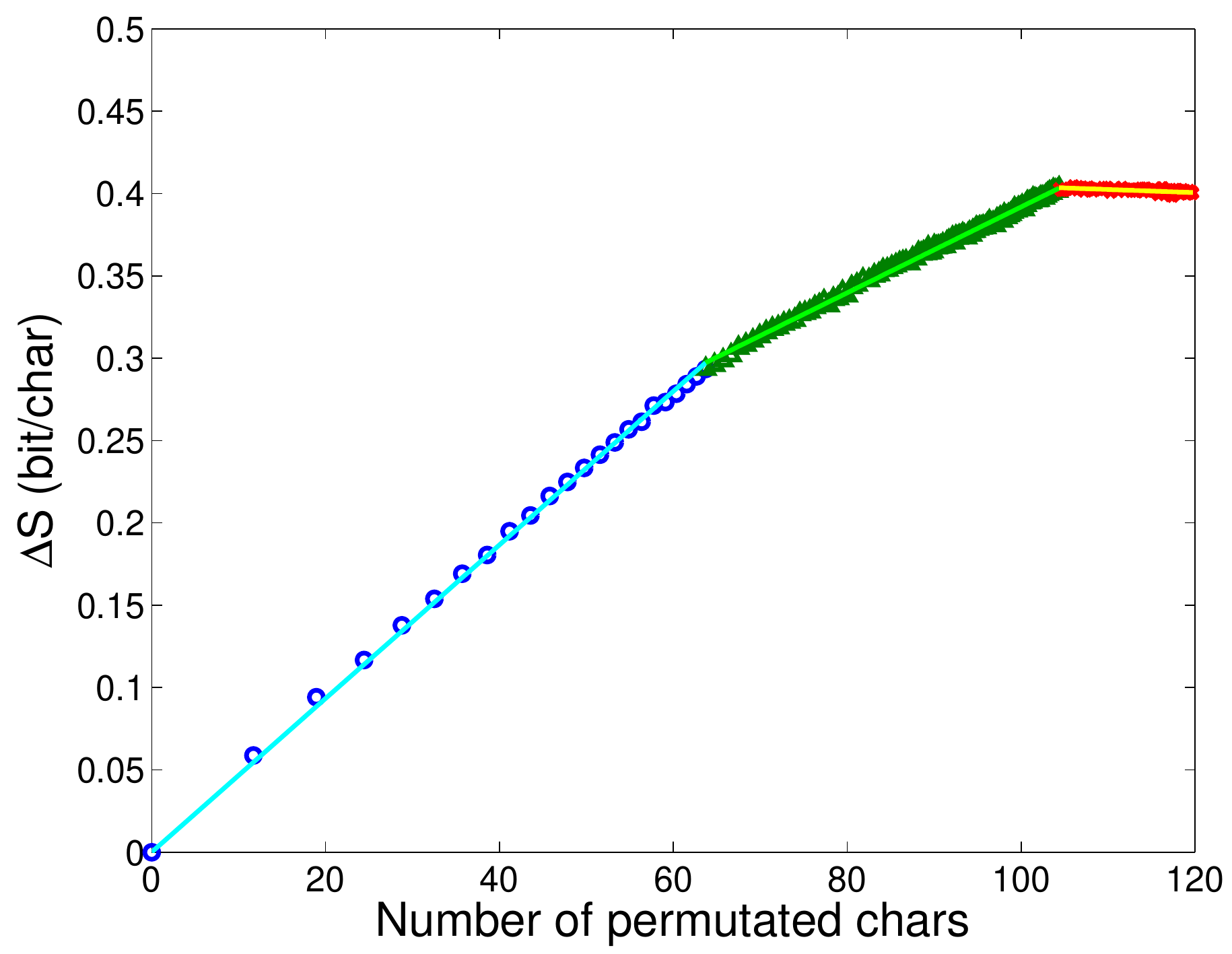}}
\caption{Performing a progressive shuffling of characters representing activities which duration is under a threshold $\Delta t'$,
we progressively break the correlations in the Jump Patterns. In this figure, we may observe how the average difference in entropy
$\Delta S$ between the shuffled and un-shuffled patterns grows with the average number of permuted chars with three different slopes.
The blue circles represent values of $\Delta t' \leq$115min, green triangles 115min$<\Delta t' \leq$12h and red diamonds 12h$ <\Delta t'$. }
\label{Svar_dt}
\end{figure}
To verify this assumption, we proceed with a different time-dependent entropic analysis. We consider a shuffling procedure the jump patterns
generated with a time frame $\Delta t_{min}=5$ minutes, by randomly permuting the activities shorter than a given value $\Delta t_s$ and comparing
the measured values of entropy $S$.
This procedure breaks the repeated sequences that the LZ algorithm finds and use for compressing the pattern, so that
the correlations between consecutive groups of characters is lost and the value of $S$ increases.
In the limiting case where all the values are shuffled, the measured value of $S$ is, in principle, equal to $S_u$.
Using the shuffling procedure instead of excluding activities allows us to analyze the activity-time dependency of the correlations without changing
the length of the patterns. For this reason, it is possible to extend the range of values $\Delta t_s$ considered in the analysis.
We have computed the values of $S$ with $\Delta t_s$ spanning from 5 minutes to 24 hours. In fig.\ref{Svar_dt}, we plot
the average difference in entropy $\Delta S(\Delta t_s) = S(\Delta t_s) - S(\Delta t_{min})$ within the sample as a function of
the average number of permuted characters $N_s$ with the same $\Delta t_s$.
The greater is the slope of the curve $\Delta S(N_s)$, the faster is the variation in entropy due to the shuffling,
and therefore the stronger is the breaking of the correlations due to the shuffling procedure. It is remarkable that this curve
can be easily divided in three segments.
\footnote{Being this result obtained, for each pattern, with a fixed string length and a fixed number of different characters,
we can exclude that this result could be a consequence of some bias due to the LZ entropy estimator.}
The best fit with a multiple linear interpolation gives the temporal limits of these parts: the first part
is limited by a time frame $\Delta t_s\le 2h$, the second part by $2h< \Delta t_s\le 12h$ and the last part by $\Delta t_s> 12h$.
Therefore, when we start shuffling the activities longer than $2h$
with the shorter activities, the increase rate in the entropy measure changes suddenly (the derivative of the second segment is roughly one half
of the derivative of the first one).
Finally, when even activities longer than $12h$ begin to be included in the shuffling, the entropy variation becomes negligible.
These three behaviors in the information entropy suggests a possible classification of activities into three categories according to their
time demand: we call them respectively short, long and very long activities.
Moreover, the fact that the information introduced with shuffling is proportional to the number $N_s$ of permuted characters indicates
that each class is internally homogeneous. Each character of any value $\Delta t$ within a given class brings the same amount of information
when moved from his original position. This result is consistent with the invariance of the compression ratio changing the exclusion
time threshold $\Delta t$ plotted in fig.\ref{tpentJ} for the activities in the range 5-60 minutes.

\begin{figure}[htc!]
\centerline{\includegraphics[angle=0, width=.45\textwidth]{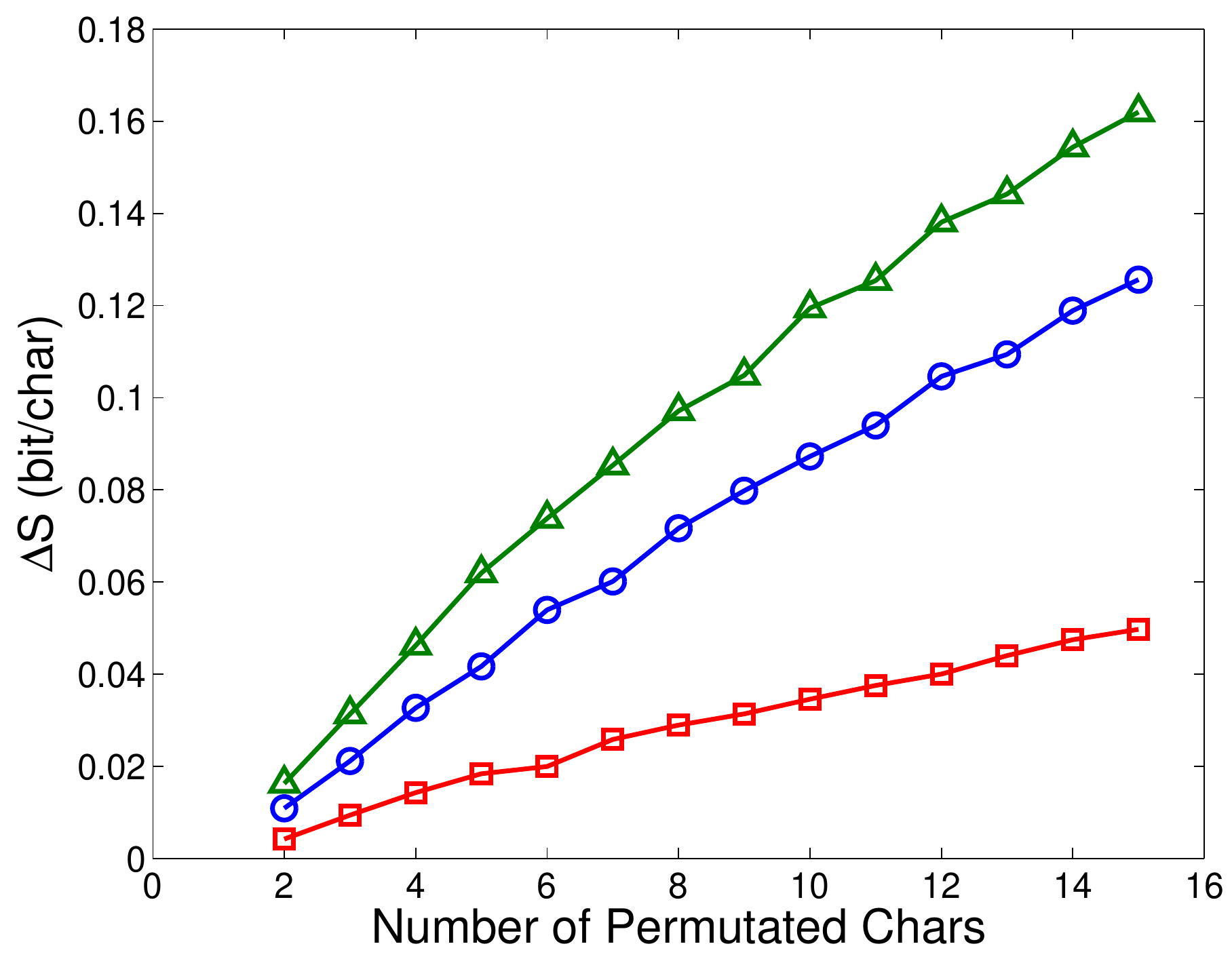}\quad\includegraphics[angle=0, width=.45\textwidth]{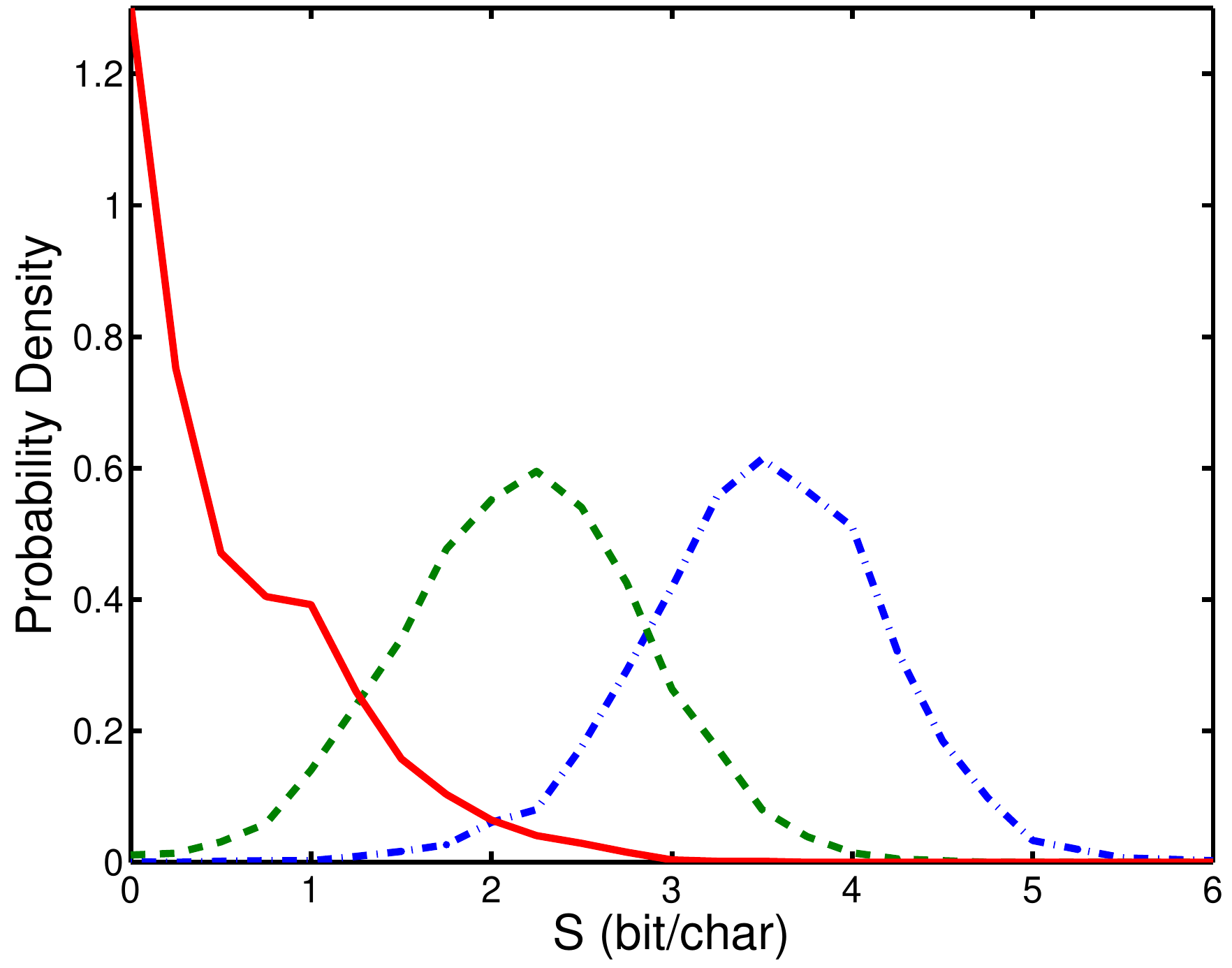}}
\caption{\textbf{(Left picture)} If we shuffle random characters belonging to the same class of activity
($\Delta t \leq$115min: blue circles , 115min$<\Delta t \leq$12h: green triangles, 12h$ < \Delta t$: red squares)
we point out that the most effective way for breaking the correlations between consecutive stops is shuffling activities of intermediate length.
\textbf{(Right picture)} The distribution of the Shannon entropies given by the frequencies of the characters in the
different classes ($\Delta t \leq$115min:  blue dot-dash line, 115min$<\Delta t \leq$12h: green dash line, 12h$ < \Delta t$:red solid line).
The short and long activities provide a Gaussian-like distribution for the Shannon entropies, but the average value of the long activity
is smaller denoting a less variability of these activities in the jump patterns.
The very long activities tend to have a less spread distribution of frequencies (the peak at $S=0$ for the activities over $12h$ indicates that
frequently only one location (probably home) is visited for such a long time.}
\label{SUint}
\end{figure}

To avoid any consequence of the progressive shuffling with a threshold going from short to long activities, we have analyzed them separately.
We have shuffled a progressive number $N_s$ of characters picked at random within each group, and we measure again the variations in the information entropy.
This analysis has been performed on a set of $\approx 3800$ jump patterns where each individual has performed at least 15 activities (not necessarily different)
for each class.
The results, shown in fig.\ref{SUint} (left), point out that shuffling the characters representing long activities increases the entropy
faster than shuffling the characters of short activities. The very long activities have the smaller derivative,
as we could have expected, as only a few locations are visited for such a long time, thus we often shuffle identical characters.
Comparing the figures \ref{SUint} (left) and \ref{Svar_dt} we remark that
if we shuffle first the short activities and then the long activities, the rate $\Delta S/N_s$ relative to the long activity
part is smaller than the rate relative to the short activities, whereas if we start shuffling the long activities, the rate is bigger.
Besides, if we proceed with the shuffling of the very long activities after having shuffled the short and the long, the shuffling is
ineffective, while the shuffling inside the class have a finite increasing rate for the entropy.
Therefore the structure of mobility patterns involving long activities seems to contain more information than that of the short and very long activities, the last strongly mainly representing the circadian rhythm that introduces periodic structures in the jump patterns.
This interpretation is confirmed by the Shannon entropy distributions calculated using the frequencies of characters of the three
classes (see fig.\ref{SUint} right). The curve representing the long activities has its values concentrated under 1 bit/char, with a peak at zero,
that means that we have usually only one character (probably the overnight stops at home). The comparison of distributions for the short and
long activities points out that the variability in the choice of short activities is larger than that of long
activities, since the $S_u$ average value of the last ones is smaller. This suggests that the difference slopes in the entropy increase due to the shuffling
of these classes (fig.\ref{SUint} left) is very significative, as the long activities have a faster rate of entropy growth with the character permutations,
even if it is more probable that these permutations are ineffective because of the repetition of the same character.
The previous results suggest the existence of a
hierarchical structure in the mobility patterns. This hierarchy appears in the repeated sequences that the LZ algorithm recognizes for the compression.
Those repeated sequences, representing individual habits, are more easily broken if we shuffle the long stops than if we shuffle the short ones.
But, if we shuffle first the short activities, shuffling then the long activities becomes less efficient in breaking these sequences.
In our opinion, this can be explained by the assumption that a significative part of the repeated sequences are constituted by clusters of
activities where a long activity plays a pivoting role between the short activities. Shuffling the long activity breaks more efficiently those sequences,
whereas if we first shuffle all the short activities, then shuffling the long activities has not strong consequences. 
The very long ($>12h$) activities are playing the role of central nodes, which act as cornerstones of the mobility schedule.
Around them repeated clusters are formed, dominated by long activities, which may correspond to daily tours\cite{schneider2013} or part of them.
We may remark that this interpretation is consistent with the time usage model where the daily schedule is created progressively, starting from the activity with the long duration
and progressively using the time left in the timetable\cite{bazzani2010}.

\section{A simple Markov model on individual mobility networks}

\begin{figure}[htc!]
\centerline{\includegraphics[angle=0, width=.45\textwidth]{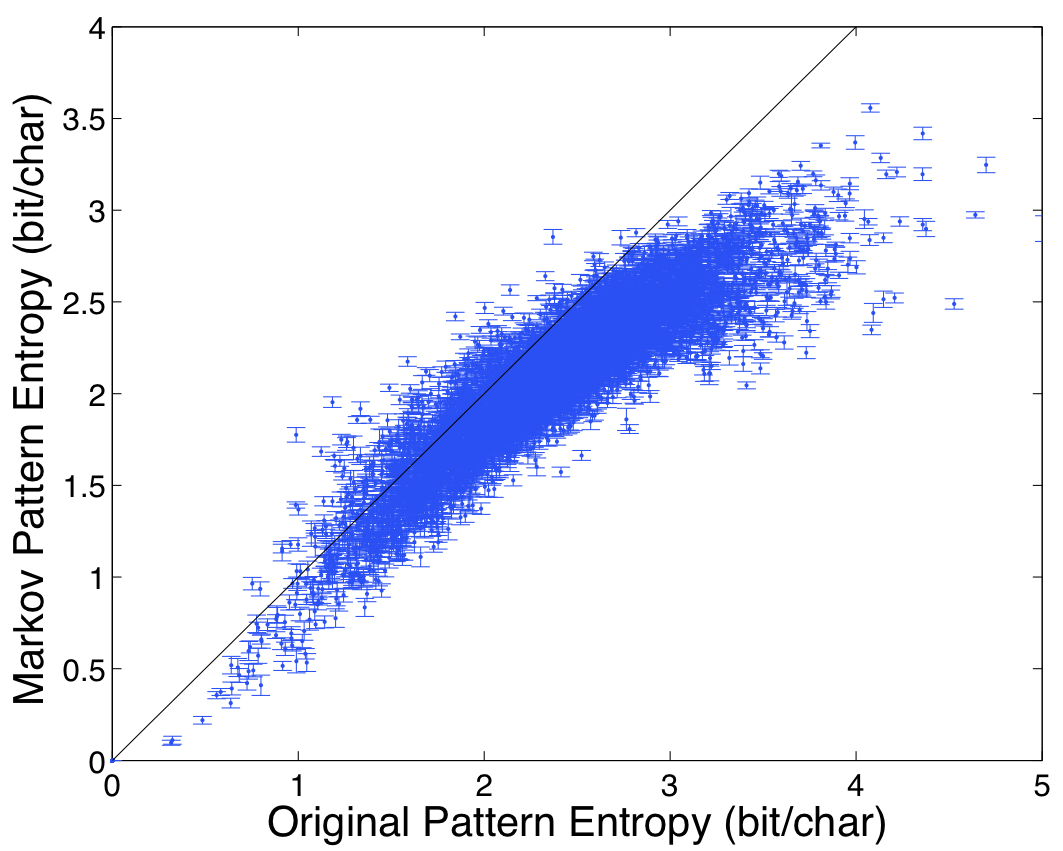}\quad
\includegraphics[angle=0, width=.45\textwidth]{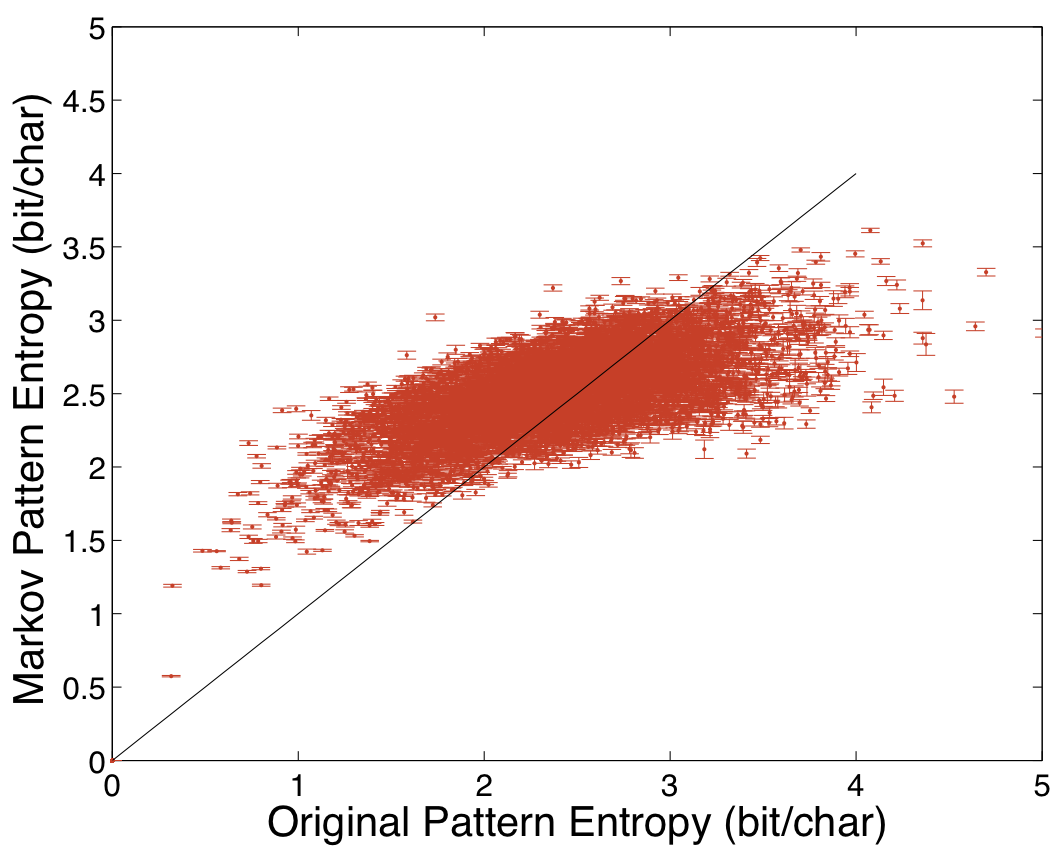}}
\caption{
Relationship between empirical entropies of jump patterns and entropies of patterns generated by a Markov process with transition probabilities $p^w_{i\to j}$
(Left) and $p^t_{i\to j} $(Right) (see eqs. (\ref{adiaprobmatrix}) and (\ref{weigthProbmatrix})). In both pictures the solid black line represents the identity.
}
\label{markovS}
\end{figure}

\begin{figure}[htc!]
\centerline{\includegraphics[angle=0, width=.45\textwidth]{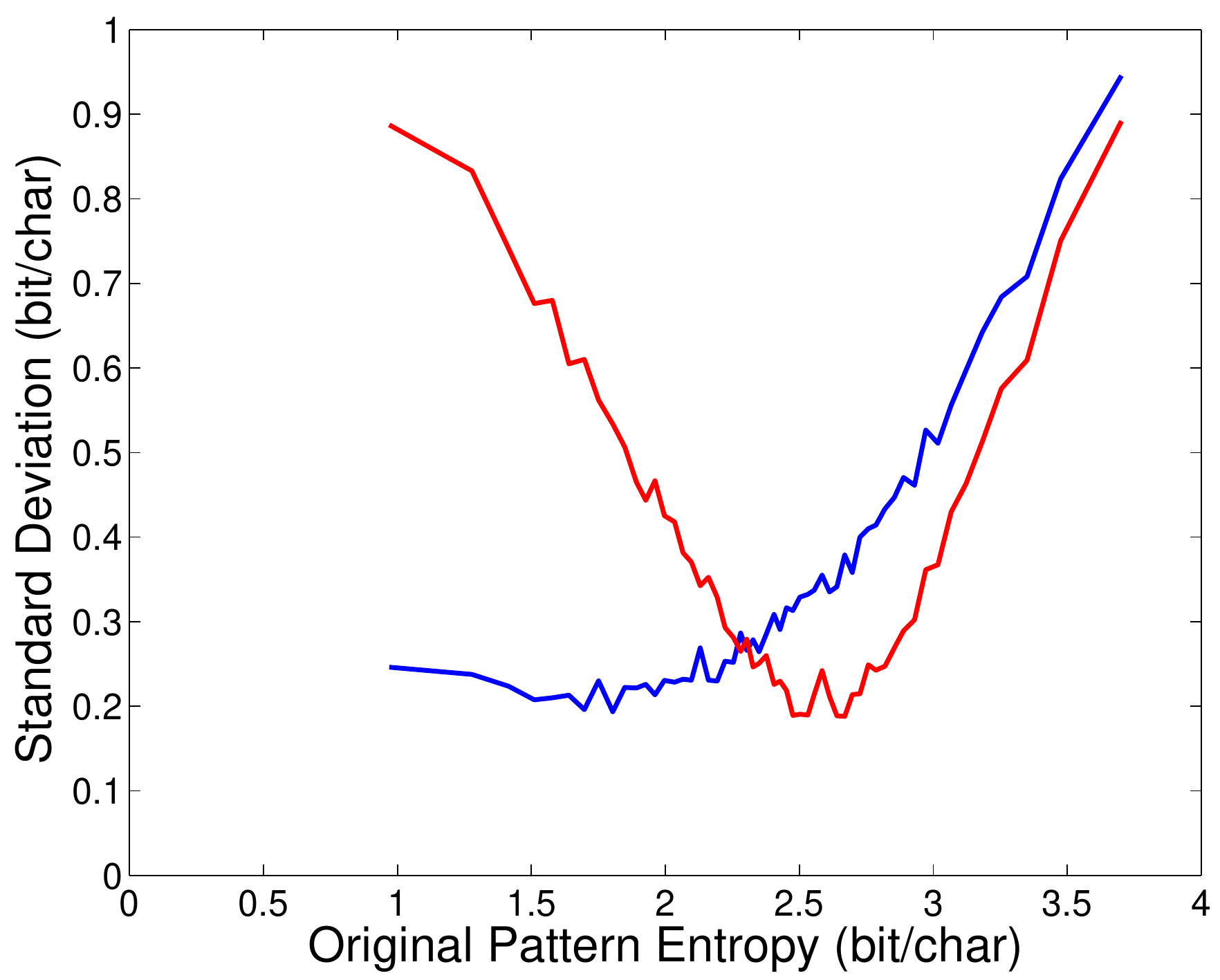}\quad\includegraphics[angle=0, width=.45\textwidth]{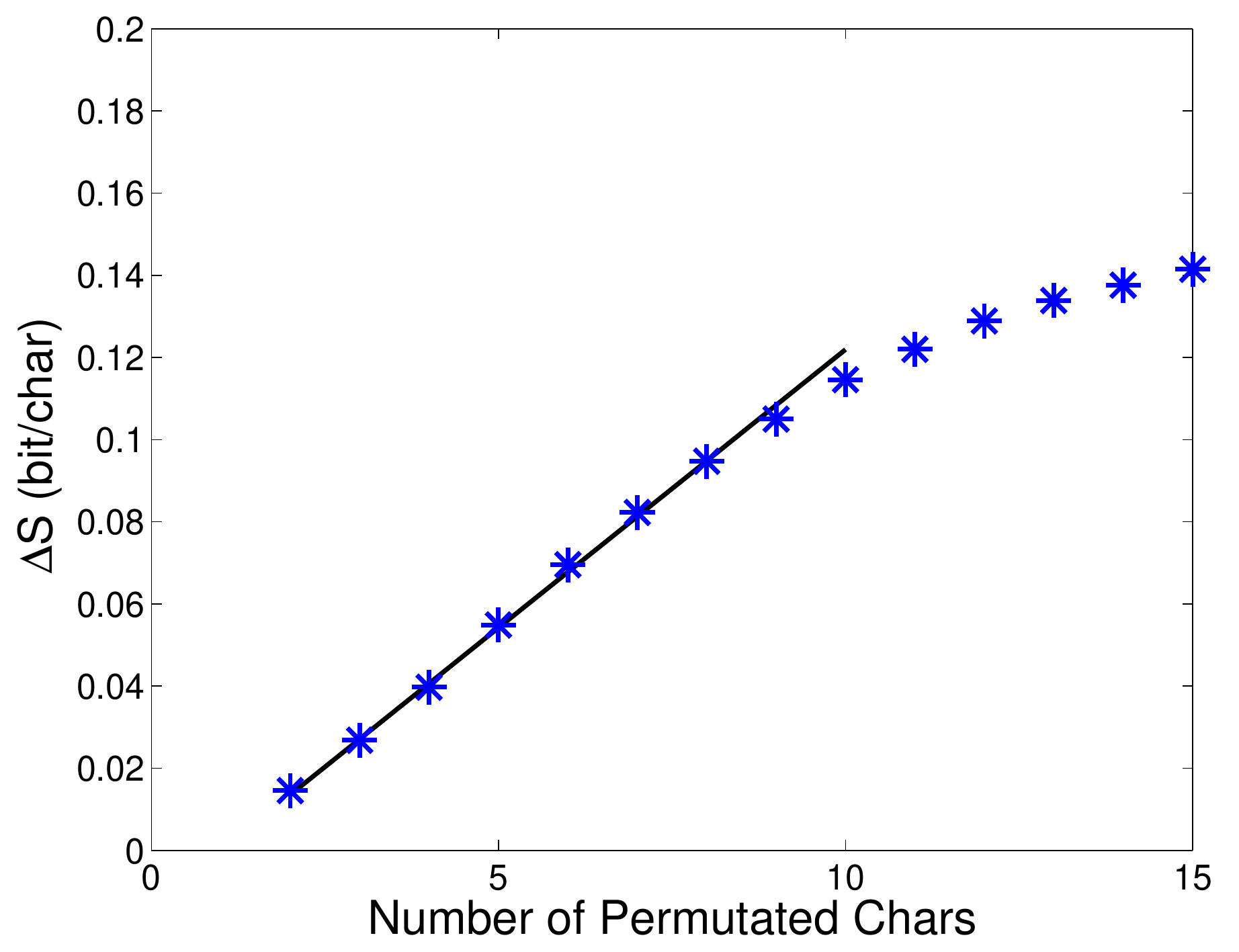}}
\caption{(Left picture) Standard deviation of the generated pattern entropies from the empirical ones. The blue curve represents deviations of
patterns generated with  $p^w_{i\to j}$ while the red one represents deviations of patterns generated with $p^t_{i\to j}$.
From this graph emerges that the firsts patterns are in good correspondence for a much wider range of values of entropies,
while in case of highly informative patterns both processes fail to emulate reality. (Right picture)
Effect of progressive rewiring of the individual networks in the values of pattern entropies generated by the transition probabilities
(\ref{weigthProbmatrix}). For a small number of rewired links, the variation grows linearly, as seen for the empirical patterns in fig.\ref{SUint}.
}
\label{markovComparison}
\end{figure}

We finally evaluate if the information of the observed jump patterns can be successfully modeled with discrete-time Markov
processes on individual mobility networks. To each individual we associated a mobility network whose nodes are the different locations
linked according to the performed trips. We have considered both from a topological point of view using the undirected adjacency matrix $a^u_{ij}$ and
introducing a weight proportional to the number of performed trips for each link $w^u_{ij}$.
Then we have computed the information entropies of patterns generated by the Markov process, whose transition probability matrix is computed from the
network adjacency matrix
\begin{equation}
p^t_{i\to j} = \frac{a^u_{ij}} {\sum_k a^u_{ik}}
\label{adiaprobmatrix}
\end{equation}
or from the weight matrix
\begin{equation}
p^w_{i\to j} = \frac{w^u_{ij}} {\sum_k w^u_{ik}}
\label{weigthProbmatrix}
\end{equation}
If the first process is able to reproduce the observed information entropy, then it is the topological structure of the mobility network
that carries the information to describe the individual mobility. On the contrary one has to take into account decision strategies
of individuals that are represented by the weights $w^u_{ij}$.
For each individual network, 10 patterns of 2000 characters have been generated to reducing both the statistical
error and the systematic error due to the entropy estimator. The values of entropy found are represented in figure \ref{markovS} against
the entropies measured for the empirical patterns.
The results show that for low-entropy value, the Markov process
with the transition matrix (\ref{weigthProbmatrix}) generates patterns considerably more similar to empirical ones whereas the
patterns of the Markov process based on the transition matrix (\ref{adiaprobmatrix}) show a systematic deviation.
However when one considers highly informative patterns, both cases present significant differences from the empirical entropy measure.
These difference can be caused by the limited length of the empirical
mobility patterns, that is probably insufficient for the convergence
of the LZ estimator to the real value of $S$ when the number of different characters is high. In the figure
\ref{markovComparison} (left) we plot the standard error of the two Markov model as a function of the information entropy of the mobility patterns.

The numerical results confirm that the Markov process defined by the transition matrix (\ref{weigthProbmatrix})
can be used for explaining the observed entropy measures for individual mobility patterns in a wide range of values.
Finally, using this model we have introduced a progressive rewiring in the
multigraph\cite{newman2003} by exchanging the trips between the nodes and redefining the weights $w^u_{ij}$, in order to compare
the entropy increase with the results of the shuffling on the mobility patterns (see previous section).
The simulations plotted in fig.\ref{markovComparison}(right) show an initial linear trend with a slope (0.014 bit/char) that lies between
the empirical slopes measured for short activities ($\Delta t_s<2h$ (0.010 bit/char) and long activities ($2h<\Delta t_s <12 h$ (0.015 bit/char))
(see fig.\ref{SUint} left).
This result verifies that
the information carried by the weighted mobility network is sufficient for describing the information
entropy of the dynamical process that has generated it, but it does not take into account the hierarchical structures
in the patterns according to the activities duration, that are suggested by the time-shuffling study of empirical patterns.

\section{Conclusions}

The analysis of entropy measures of mobility patterns from GPS signals in urban context (Florence), points out that it is possible to identify three different
classes of activities: short activities $\Delta t <$ 2h, long
activities 2h $<\Delta t <$ 12h and very long activities 12h
$<\Delta t$. Each of these categories is homogeneous and takes part
in different ways at the structure of the individual mobility
patterns. The homogeneity within the short activities is confirmed
by the constancy of the compression ratio observed in the jump
pattern analysis. Very long activities can be identified with
the overnight stop at home, together with only a few other un-frequently
taken alternatives. This daily return to the own house constitutes
the base structure of the individual mobility pattern. Short and long
activities are performed according to the circadian rhythm, but among them
exists a hierarchical relationship, where long activities tend
to plays a more central role than short activities.
Long activities are most likely planned before
short activities, thus playing a pivotal role for them.
In fact, short activities arrange themselves around the long (or very long)
activities, to form sequences that are then repeated in
different days\cite{schneider2013}.
However, on the contrary to what one expected, short activities
do not seem be executed randomly. Indeed, the homogeneity in the
classes indicates that even the shortest ones are eventually able to
concur in a systematic mobility, pointed out by the formation of
repeated sequences.
A recent study\cite{bagrow2012} proposes that the spatiotemporal
structure of human mobility patterns can be flow-wise partitioned in
groups of related nodes called habitats, and that those habitats
tend to be more spatially cohesive than the total mobility. This
suggests that the repeated sequences may reflect this habitat
structure and thus be geographically related. 
The individuals have
to optimize their daily schedule, as their mobility limited by the
mobility energy\cite{kolbl2003}\cite{gallotti2012}, and therefore activities that
can be performed in near locations are most likely to be made in
chain.
The reason for the threshold the values 2 hours and 12 hours requires additional information. Probably the value of 12 hours is related to the circadian
rhythm, whereas the value of 2 hours (or, more exactly, 115 minutes) may
instead be a duration characteristic of the Florence area.
Finally we have shown that the information entropy measures of the mobility patterns can be reproduced
by a Markov process whose transition probability matrix is computed using the empirical weights of the
undirected individual mobility networks.\par\noindent Further
analysis on activity patterns of individuals living in different cities might lead to a better understanding and a generalization of these results.

\section{Acknowledgements}

This work has been partially supported by the PRIN-project 2008T3HFT9\_002 (MIUR, Italy). We thank Octo Telematics S.p.A. for the access to the GPS data base.

\end{document}